\documentclass[preprint,aps,prd,groupedaddress,showpacs,nofootinbib]{revtex4}
\usepackage{graphicx}
\usepackage{amsmath}
\usepackage{bm}
\usepackage{epsfig}

\begin{document}

\preprint{ANL-HEP-PR-06-57}
\title{\mbox{}\\[10pt]
Exclusive Two-Vector-Meson Production
from $\bm{e^+ e^-}$ Annihilation}


\author{Geoffrey T. Bodwin}
\affiliation{
High Energy Physics Division,
Argonne National Laboratory,
9700 South Cass Avenue, Argonne, Illinois 60439}

\author{Eric Braaten}
\affiliation{
Physics Department, Ohio State University, Columbus, Ohio 43210}

\author{Jungil Lee and Chaehyun Yu}
\affiliation{
Department of Physics, Korea University,
Seoul 136-701, Korea
}


\date{\today}
\begin{abstract}
The exclusive production of pairs of vector mesons with $J^{PC}  =
1^{--}$ in $e^+ e^-$ collisions can proceed through $e^+ e^-$
annihilation into two virtual photons. At energies much greater than
the meson masses, the cross section is dominated by the independent
fragmentation of the virtual photons into the vector mesons. The
fragmentation approximation is used to calculate the cross sections and
angular distributions for pairs of vector mesons that can be produced at
the $B$ factories. The predicted cross sections for $\rho^0 + \rho^0$ and
$\rho^0 + \phi$ production agree with recent measurements by the BaBar
Collaboration. For the production of two charmonium vector mesons, the
nonfragmentation corrections to the cross sections are calculated by
using the NRQCD factorization formalism. The predicted cross sections
for $J/\psi + J/\psi$ and $J/\psi + \psi(2S)$ production are compatible
with upper limits set by the Belle Collaboration.
\end{abstract}


\maketitle


\section{Introduction
 \label{intro}}


The production of hadrons in $e^+ e^-$ collisions 
at energies well below the mass of the $Z^0$ proceeds 
predominantly through the annihilation of $e^+ e^-$ into a 
virtual photon.  This process can only produce final states 
with charge conjugation quantum number $C=-1$.
Final states with $C=+1$   can be produced by the annihilation 
of $e^+ e^-$ into two virtual photons, but the cross sections 
for these processes are suppressed by a factor of
$\alpha^2$, where $\alpha \simeq 1/137$ is the QED fine 
structure constant.  Final states with $C=+1$   can also be produced 
by the annihilation of $e^+ e^-$ into a virtual $Z^0$, 
but the cross sections for these processes are suppressed by 
a factor of $(s/M_Z^2)^2$ if the $e^+ e^-$ center-of-mass energy 
$\sqrt{s}$ is small compared to $M_Z$.  Thus, unless there are 
enhancement factors to compensate for the factors   of $\alpha^2$
or $(s/M_Z^2)^2$, the production   cross sections at $B$-factory 
energies   for $C=+1$   states 
are orders of magnitude smaller than those for $C=-1$   states.

In $e^+ e^-$ collisions with large center-of-mass energy $\sqrt{s}$,
factors of  $\sqrt{s}/m$, where $m$ is a hadronic mass scale,
can compensate for the suppression factor of $\alpha^2$. 
An example of such
compensation can be seen in the exclusive double-charmonium
production process $e^+ e^- \to J/\psi + J/\psi$
\cite{Bodwin:2002fk,Bodwin:2002kk}. At the beam energy of the $B$
factories, $\sqrt{s}/2=5.29$ GeV, the 
leading-order prediction for the production cross section for
the $C=+1$ final state $J/\psi + J/\psi$ has roughly the same order of
magnitude as that for the corresponding $C=-1$ final state $J/\psi +
\eta_c$~\cite{Braaten:2002fi}. The production cross section for
$J/\psi + \eta_c$ scales as $\alpha^2 \alpha_s^2 m_c^6/s^4$, where
$m_c$ is the charm-quark mass. In contrast, the production cross
section for $J/\psi + J/\psi$ scales as $\alpha^4/s$ because there is a
photon-fragmentation contribution that corresponds to the annihilation
process $e^+ e^- \to \gamma^* + \gamma^*$, followed by independent
fragmentation processes $\gamma^* \to J/\psi$. The enhancement
factor of $(s/m_c^2)^3$ in the production cross section for $J/\psi +
J/\psi$ relative to that for $J/\psi + \eta_c$ partially compensates 
for the suppression factor of $(\alpha/\alpha_s)^2$.  

A similar enhancement mechanism is present   in the exclusive
production of a pair of light vector mesons $V_1 + V_2$,   both of
which have quantum numbers $J^{PC} = 1^{--}$.  The production process
proceeds   predominantly through the annihilation process $e^+ e^- \to
\gamma^* + \gamma^*$,   followed by the   independent fragmentation
processes $\gamma^* \to V_i$. We can compare the production cross
section for $V_1 + V_2$ to that for $V + P$, where $P$ is a light
pseudoscalar meson. In the ratio of the former cross section to the
latter, there is an enhancement factor of $(s/\Lambda_{\rm QCD}^2)^3$
that compensates for the suppression factor of $(\alpha/\alpha_s)^2$.  
In this paper, we calculate the cross sections and angular distributions
for the exclusive production of pairs of vector mesons with $J^{PC}  =
1^{--}$.

The remainder of this paper is organized as follows. In Sec.~\ref{ampl},
we discuss the various contributions to the amplitudes for the
production of two vector mesons and estimate their relative sizes. In
Sec.~\ref{sec:differential}, we give expressions for the differential
cross sections for the production of two vector mesons in the
fragmentation approximation and give numerical results for the cross
sections. Sec.~\ref{sec:nonfrag} contains a computation of the
nonfragmentation corrections to the cross sections for the production of
two charmonium vector mesons. In Sec.~\ref{sec:comparison}, we compare
our results with experimental measurements and with previous theoretical
calculations. The Appendix contains the expressions for the
nonfragmentation corrections to the differential cross sections for two
charmonium vector mesons.
\section{Production amplitudes
 \label{ampl}}

\begin{figure}
\begin{tabular}{lcr}
\epsfig{file=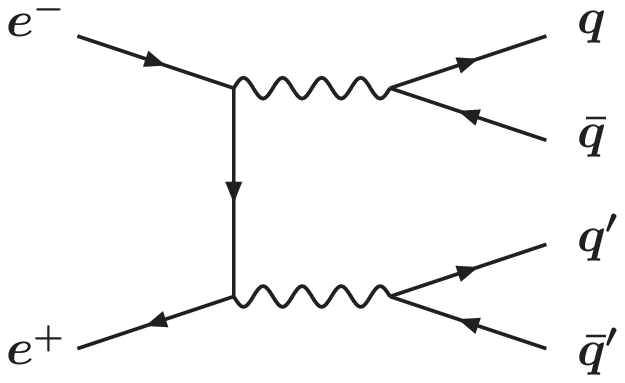,width=7cm}&\qquad\qquad\qquad&
\epsfig{file=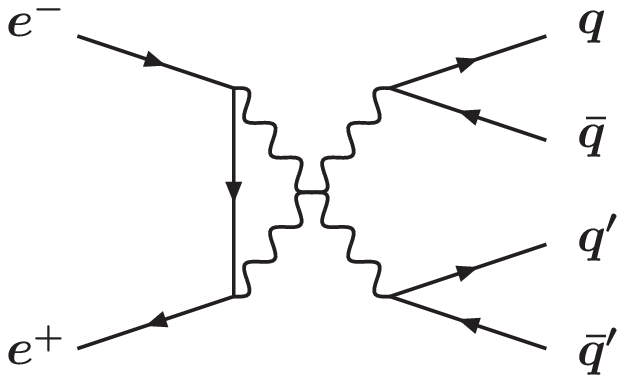,width=7cm}
\end{tabular}
\vspace*{8pt}
\caption{\label{fig1}%
QED fragmentation diagrams that contribute to exclusive 
two-vector-meson production in $e^+ e^-$ collisions if the two 
vector mesons have quark contents $q \bar q$ and $q' \bar q'$. 
}
\end{figure}

\begin{figure}
\begin{tabular}{lcr}
\epsfig{file=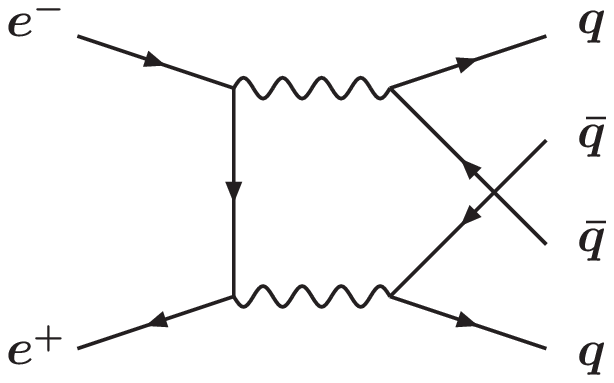,width=7cm}&\qquad\qquad\qquad&
\epsfig{file=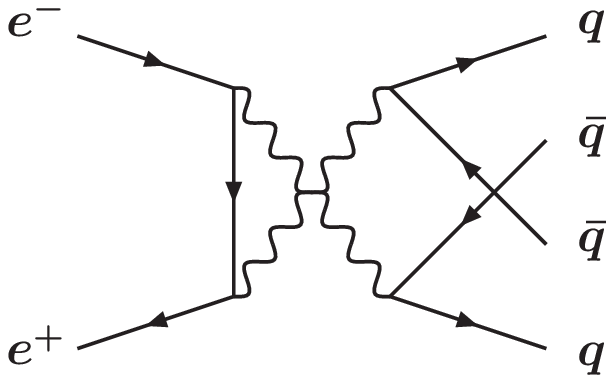,width=7cm}
\end{tabular}
\vspace*{8pt}
\caption{\label{fig2}%
QED nonfragmentation diagrams that contribute to exclusive 
two-vector-meson production in $e^+ e^-$ collisions if the two 
vector mesons have the same quark content $q \bar q$. }
\end{figure}

The lowest-order   QED diagrams that correspond to the creation of the
quarks in the two vector mesons $V_1$ and $V_2$ are shown in
Figs.~\ref{fig1} and \ref{fig2}. If the two vector mesons have different
quark contents $q \bar q$ and $q' \bar q'$, then only the diagrams in
Fig.~\ref{fig1} are present.  Once the quarks are created, the formation
of the vector mesons involves the exchange of arbitrarily many gluons
between the quark lines.  Gluons can be exchanged between the quark and
antiquark that form a given meson or between quarks or antiquarks that
form different mesons.  However, factorization theorems and the
scaling of various contributions yield a simplification of this picture
in the limit $E_{\rm beam}\to \infty$. In what follows, we take the beam
energy $E_{\rm beam}= \sqrt{s}/2$ to be much greater than the masses
$m_{V_i}$ of the vector mesons, which implies that $E_{\rm beam}$ is
also much greater than $\Lambda_{\rm QCD}$.

Let us first consider the exchange of gluons only between the $q$ and
$\bar q$ within a given meson. Then, the diagrams in
Fig.~\ref{fig2} are suppressed in comparison to those in
Fig.~\ref{fig1}. In the diagrams of Fig.~\ref{fig1}, the virtual-photon
propagators are $1/m_{V_i}^2$, and the two hadronic electromagnetic
currents each give a factor of order $\Lambda_{\rm QCD}$ for light
mesons and $m_c$ for charmonium mesons.  In the diagrams of
Fig.~\ref{fig2}, the virtual-photon propagators are of order $1/E_{\rm
beam}^2$, and the two hadronic electromagnetic currents each give factors
of order $E_{\rm beam}$. This leads to the following suppression factors
for the diagrams of Fig.~\ref{fig2} relative to those of
Fig.~\ref{fig1}: $\Lambda_{\rm QCD}^2/ E_{\rm beam}^2$ for two light
mesons and $m_c^2/E_{\rm beam}^2$ for two charmonium mesons. We have
taken the light-meson masses to be of order $\Lambda_{\rm QCD}$ and the
charmonium masses to be of order $m_c$. (The diagrams of Fig.~\ref{fig2}
do not contribute in the case of a light meson and a charmonium meson.)

Next, let us consider the exchange of gluons between quarks or
antiquarks in different mesons, with some of those gluons soft or, in
the case of a light meson, collinear. In the case of two light mesons,
one can use standard methods for proving factorization theorems
\cite{Collins:1989gx} to show that the soft contributions cancel up to
corrections of relative order $\Lambda_{\rm QCD}^4/E_{\rm beam}^4$ and
the collinear contributions cancel up to corrections of relative order
$\Lambda_{\rm QCD}^2/E_{\rm beam}^2$.\footnote{The results that we quote
here apply to the case in which one sums over all polarizations of the
final-state mesons. If one observes the polarization of one meson, then
the soft contributions cancel up to terms of relative order
$\Lambda_{\rm QCD}^3/E_{\rm beam}^3$, and the collinear contributions
cancel up to terms of relative order $\Lambda_{\rm QCD}/E_{\rm beam}$.
If one observes the polarization of both mesons, then the soft
contributions cancel up to corrections of relative order $\Lambda_{\rm
QCD}^2/E_{\rm beam}^2$ and the collinear contributions cancel up to
corrections of relative order $\Lambda_{\rm QCD}/E_{\rm beam}$. Similar
reductions of the suppression factors for soft and collinear
contributions occur in the cases in which one or more charmonium mesons
are produced.} Factorization theorems are not well established for the
production of heavy-quarkonium mesons. However, on the basis of existing
factorization technology, it seems plausible, in the case of one light
meson and one charmonium meson, that soft contributions cancel up to
terms of relative order $\Lambda_{\rm QCD}^2(m_cv)^2/E_{\rm beam}^4$ and
that collinear contributions cancel up to terms of relative order
$\Lambda_{\rm QCD}^2/E_{\rm beam}^2$. Here $v$ is the typical velocity
of the charm quark or antiquark in the charmonium rest frame.
($v^2\approx 0.3$.) In the case of two charmonium mesons, there are no
collinear singularities. It is plausible that soft contributions cancel
up to terms of relative order $(m_cv)^4/E_{\rm beam}^4$ in this case.

Finally, let us consider the exchange of hard gluons between quarks or
antiquarks in different mesons. (Hard gluons have energies and momenta
of order $E_{\rm beam}$.) In the case of the diagrams of
Fig.~\ref{fig2}, exchanges of hard gluons are suppressed as
$\alpha_s(E_{\rm beam})$. In the case of the diagrams of Fig.~\ref{fig1}, 
it is necessary to exchange at least two gluons between the mesons 
in order to keep the $q \bar q$ pair in a color-singlet state. Hence, 
such exchanges are suppressed as $\alpha_s^2(E_{\rm beam})$. They are 
also suppressed by a kinematic factor that arises as follows. In the
diagrams of Fig.~\ref{fig1} without hard-gluon exchange, the
virtual-photon propagators are $1/m_{V_i}^2$ and the two hadronic
electromagnetic currents each give a factor of order $\Lambda_{\rm QCD}$
for light mesons and $m_c$ for charmonium mesons. In the diagrams of
Fig.~\ref{fig1} with hard-gluon exchanges, the virtual-photon
propagators are of order $1/E_{\rm beam}^2$, and the two hadronic
electromagnetic currents each give factors of order $E_{\rm beam}$. This
leads to the following additional suppression factors for hard-gluon
exchanges in the diagrams of Fig.~\ref{fig1}: $\Lambda_{\rm QCD}^2/
E_{\rm beam}^2$ for two light mesons, $m_c\Lambda_{\rm QCD}/ E_{\rm
beam}^2$ for a light meson and a charmonium meson, and $m_c^2/E_{\rm
beam}^2$ for two charmonium mesons. Again, we have taken the light-meson
masses to be of order $\Lambda_{\rm QCD}$ and the charmonium masses to
be of order $m_c$.

Let us now summarize the results of the preceding analyses.  We conclude
that we need only consider the fragmentation diagrams in
Fig.~\ref{fig1}, up to terms of relative order $\Lambda_{\rm QCD}^2/
E_{\rm beam}^2$ for two light mesons and $m_c^2/E_{\rm beam}^2$ for two
charmonium mesons. (The diagrams of Fig.~\ref{fig2} do not contribute in
the case of a light meson and a charmonium meson.) We also conclude that
we need only consider the QCD corrections to the diagrams of
Fig.~\ref{fig1} that correspond to the exchange of gluons between the
$q$ and $\bar q$ in each meson, up to terms of relative order
$\Lambda_{\rm QCD}^2/E_{\rm beam}^2$ for two light mesons, $\Lambda_{\rm
QCD}^2/E_{\rm beam}^2$ or $\alpha_s^2(E_{\rm beam}) \, (m_c\Lambda_{\rm
QCD}/ E_{\rm beam}^2)$ for a light meson and a charmonium meson, and
$(m_cv)^4/E_{\rm beam}^4$ or $\alpha_s^2(E_{\rm beam}) \, (m_c^2/E_{\rm
beam}^2)$ for two charmonium mesons.\footnote{
There are several recent papers in which the rate for the process
$\gamma^* \gamma^* \to \rho^0 \rho^0$ is calculated in the high-energy
and diffractive limit $s \gg -t$, where $s$ and $t$ are the
Mandelstam variables 
\cite{Pire:2005ic,Enberg:2005eq,Ivanov:2005gn,Pire:2006ik}. 
In this limit, the dominant
contributions to this process involve the exchange of gluons between the
quarks and antiquarks that form different mesons.  There are analogous
contributions to the process $e^+ e^- \to \rho^0 \rho^0$, since it also
proceeds through $\gamma^* \gamma^* \to \rho^0 \rho^0$. However, as we
have discussed above, these contributions are suppressed, and the
dominant mechanism in $e^+ e^- \to \rho^0 \rho^0$ is independent photon
fragmentation. 
Independent photon fragmentation is not considered in
Refs.~\cite{Pire:2005ic,Enberg:2005eq,Ivanov:2005gn,Pire:2006ik}
because it corresponds in the process
$\gamma^* \gamma^* \to \rho^0 \rho^0$ to a disconnected diagram.
}

The QCD corrections that involve the exchange of gluons between
the $q$ and $\bar q$ in each meson are precisely those that correspond
to the formation of a vector meson $V$ from a $q \bar q$ pair that has
been created at a point. They can be expressed in terms of a QCD
matrix element of the electromagnetic current:
\begin{equation}
J^\mu(x) = e_u\bar u(x) \gamma^\mu u(x) 
+ e_d\bar d(x) \gamma^\mu d(x) 
+ e_s\bar s(x) \gamma^\mu s(x) 
+ e_c\bar c(x) \gamma^\mu c(x),
\label{current}
\end{equation}
where $e_u = e_c = +\frac{2}{3}$ and $e_d = e_s = -\frac{1}{3}$.
The matrix element can be used to define a vector-meson-photon
coupling constant $g_{V \gamma}$:
\begin{equation}
\langle V(\lambda) | J^\mu(x=0) | 0 \rangle = 
g_{V \gamma} \, \epsilon^\mu(\lambda)^*,
\label{GV-def}
\end{equation}
where $\lambda$ is the helicity of the vector meson.
The coupling constant $g_{V\gamma}$ can be determined from the 
electronic width of the vector meson:
\begin{equation}
\Gamma[V \to e^+ e^-] = \frac{4\pi\alpha^2 g^2_{V\gamma}}{3m_V^3}.
\label{gamma-v-ee}
\end{equation}
Using the measured electronic widths that are given in Ref.~\cite{PDG2006},
we obtain the values of $g_{V\gamma}$ that are shown in Table~\ref{constants}.
The stated uncertainties arise 
from the electronic widths. The uncertainties in the masses are at 
most $0.06\%$ and can be neglected.  
\begin{table}[t]
\caption{\label{constants}%
Masses, electronic decay widths, and coupling constants $g_{V\gamma}$
for vector mesons $V$. All of the data except for those of $\rho^0$ are
taken from Ref.~\cite{PDG2006}. The data for $\rho^0$ are taken from
Ref.~\cite{Akhmetshin:2003zn} in order to maintain consistency between
the narrow-width approximation and the nonzero-width parametrization. The
uncertainties shown for $g_{V\gamma}$ are those that arise from the
uncertainties in the electronic widths of the vector mesons.}
\begin{ruledtabular}
\begin{tabular}{l|ccccc}
$V$ & Mass~(MeV) & Width~(keV) & $g_{V\gamma}$~(GeV$^2$)\\
\hline
$\rho^0$&
$ 775.65  \pm 0.64 \pm 0.50  $& $7.06\pm 0.11 \pm 0.05$& 
$0.122 \pm 0.001$
\\
$\omega$&
$ 782.65 \pm 0.12 $& $0.60\pm 0.02$& $0.036 \pm 0.001$
\\
$\phi$&
$1019.46 \pm 0.019$& $1.27\pm 0.04$& $0.078 \pm 0.001$
\\
$J/\psi$&
$3096.916\pm 0.011$& $5.55\pm 0.14$& $0.860 \pm 0.011$
\\
$\psi(2S)$&
$3686.093\pm 0.034$& $2.48\pm 0.06$& $0.746 \pm 0.009$
\end{tabular}
\end{ruledtabular}
\end{table}

\section{Cross sections in the fragmentation 
approximation}\label{sec:differential}

We express our results for the cross section for $e^+e^-\to V_1+V_2$ 
in terms of dimensionless 
variables $r_{V_1}$ and $r_{V_2}$, which are defined by
\begin{eqnarray}
r_V &=& \frac{ m_V }{ \sqrt{s}}.
\label{r-def}
\end{eqnarray}
If we take $\sqrt{s}$ to be $10.58$ GeV, the center-of-mass energy
of the $B$ factories, then $r_V^2$ is small for charmonium
($r_{J/\psi}^2= 0.086$) and very small for light vector mesons
($r_\rho^2 = 0.0054$). The differential cross sections
$d\sigma/d\cos\theta$ for each of the helicity states of $V_1$ and $V_2$
are
\begin{eqnarray}
\frac{d\sigma}{dx}[V_1(\lambda_1) + V_2(\lambda_2)] &=&
\frac{16\pi^3 \alpha^4 g_{V_1\gamma}^2 g_{V_2\gamma}^2
 \lambda^{1/2}(1,r_1^2,r_2^2) F_{\lambda_1,\lambda_2}(r_1,r_2,x)}
     {s^5 \, r_1^4 r_2^4
 \big[ (1-x^2) \lambda(1,r_1^2,r_2^2) + 4r_1^2r_2^2 \big]^2},
\label{dsigmadx:VV}
\end{eqnarray}
where $x=\cos \theta$,  $r_i = r_{V_i}$, and  
\begin{eqnarray}
\lambda(a,b,c) &=& a^2+b^2+c^2-2ab -2 bc-2ca.
\end{eqnarray}
The functions $F_{\lambda_1,\lambda_2}(r_1,r_2,x)$, 
which depend on the helicities of the vector mesons, are given by
\begin{subequations}
\label{F10}
\begin{eqnarray}
F_{\pm 1,\mp 1}(r_1,r_2,x)&=& f_{+-}(r_1,r_2,x)+ f_{+-}(r_1,r_2,-x),
\label{Fpm}
\\
F_{\pm 1,0}(r_1,r_2,x)&=&f_{+0}(r_1,r_2,x)+ f_{+0}(r_1,r_2,-x),
\\
F_{0,\pm 1}(r_1,r_2,x)&=&f_{+0}(r_2,r_1,x)+ f_{+0}(r_2,r_1,-x),
\\
F_{\pm 1,\pm 1}(r_1,r_2,x)&=& f_{++}(r_1,r_2,x)+ f_{++}(r_1,r_2,-x),
\\
F_{0,0}(r_1,r_2,x)&=& 16 r_1^2 r_2^2 x^2 (1-x^2),
\end{eqnarray}
\end{subequations}%
where
\begin{subequations}
\label{ff}
\begin{eqnarray}
f_{+-}(r_1,r_2,x)&=&
\mbox{$\frac{1}{2}$}(1+x)(1-x)^3(1-r_1^2-r_2^2)^2,
\\
f_{+0}(r_1,r_2,x)&=&
r_2^2(1-x)^2[ (1-r_2^2)(1+x)-r_1^2(1-x) ]^2,
\\
f_{++}(r_1,r_2,x)&=&
\mbox{$\frac{1}{2}$}(1-x^2)[ (1+x)r_2^2(1-r_2^2)-(1-x)r_1^2(1-r_1^2) 
         + 2 r_1^2r_2^2 x ]^2.
\end{eqnarray}
\end{subequations}%
Since $r_1$ and $r_2$ are small, the cross section is largest 
if the two vector mesons are transversely polarized with opposite 
helicities: $\lambda_1 = - \lambda_2 = \pm 1$.
If the vector meson $V_i$ is longitudinally polarized,
then the cross section is suppressed as $r_i^2$.
If the vector mesons have the same helicities, then the cross section 
is suppressed as four powers of $r_1$ or $r_2$.

After summing over the helicities $\lambda_1$ and $\lambda_2$ of
the vector mesons, we obtain
\begin{eqnarray}
\frac{d\sigma}{dx}(m_{V_1},m_{V_2})&=&
\frac{32\pi^3\alpha^4 g^2_{V_1\gamma} g^2_{V_2\gamma}
      \lambda^{1/2}(1,r_1^2,r_2^2)}
      {s^5 [r_1 r_2 (1-r_1^2-r_2^2)]^4 [1-(1-\Delta)x^2]^2}
\bigg[
16 r_1^2 r_2^2 (r_1^2+r_2^2)\nonumber\\
&&\qquad +2(1-x^2)(1+r_1^4+r_2^4)\lambda(1,r_1^2,r_2^2)
-(1-x^2)^2\lambda^2(1,r_1^2,r_2^2)
\bigg],
\label{sigma-hel-sum}
\end{eqnarray}
where $\Delta = 4r_1^2 r_2^2/(1-r_1^2 -r_2^2)^2$.
The numerical value of $\Delta$ is very small for both charmonia and
light vector mesons. For example, $\Delta=0.043$ for $J/\psi + J/\psi$
and $\Delta = 0.00012$ for $\rho^0 + \rho^0$. In the arguments on the
left side of Eq.~(\ref{sigma-hel-sum}), we have indicated the
dependences on the masses of the vector mesons, as they will play a
r\^ole in a subsequent discussion of the effect of the width of the
$\rho$ meson. The total cross section $\sigma$ is obtained by
integrating over $x$ from $-1$ to $+1$ if the two vector mesons are
distinct. If the two vector mesons are identical, then the integration
range of $x$ is taken to be $0$ to $1$ in order to avoid double
counting. The integration formulas that are required  to evaluate the
definite integrals over $x$ from $-a$ to $a$ analytically are
\begin{subequations}
\label{integrals}
\begin{eqnarray}
\int_{-a}^a dx \frac{1}{[1-(1-\Delta)x^2]^2}&=& 
  \frac{a}{1-a^2+a^2\Delta}+\frac{1}{2\sqrt{1-\Delta}}
  \ln \frac{1+a\sqrt{1-\Delta}}{1-a\sqrt{1-\Delta}},
\\
\int_{-a}^a dx \frac{1-x^2}{[1-(1-\Delta)x^2]^2}&=& 
   -\frac{a\Delta}{(1-\Delta)(1-a^2+a^2\Delta)}
\nonumber\\
  &&
  +\frac{2-\Delta}{2(1-\Delta)^{3/2}}
  \ln \frac{1+a\sqrt{1-\Delta}}{1-a\sqrt{1-\Delta}},
\\
\int_{-a}^a dx \frac{(1-x^2)^2}{[1-(1-\Delta)x^2]^2}&=& 
  \frac{1}{(1-\Delta)^2}
  \left( 2a
        +\frac{a\Delta^2}{1-a^2+a^2\Delta}
  \right)
  \nonumber\\
  &&
  +\frac{\Delta(\Delta-4)}{2(1-\Delta)^{5/2}}
  \ln \frac{1+a\sqrt{1-\Delta}}{1-a\sqrt{1-\Delta}}.
\end{eqnarray}
\end{subequations}%
  
The cross sections for two vector mesons with $J^{PC} = 1^{--}$ have
also been calculated recently in Ref.~\cite{Davier:2006fu}. As was
pointed out in Ref.~\cite{Davier:2006fu}, at the level of precision of
this work, it is necessary to take into account the nonzero width of the
$\rho$ meson. Following Ref.~\cite{Davier:2006fu}, we note that
Eq.~(\ref{sigma-hel-sum}) can be generalized to the case of a continuous
mass spectrum:
\begin{equation}
\left(\frac{d\sigma}{dx}\right)_{\rm cont}
=\left(\frac{1}{4\pi^2\alpha}\right)^2\int 
d\tilde{m}_1^2 d\tilde{m}_2^2
\,\sigma_{e^+e^-\to H}(\tilde{m}_1) \,\sigma_{e^+e^-\to H}(\tilde{m}_2)
\, \frac{d\sigma_{\gamma_1^*\gamma_2^*}}{dx}(\tilde{m}_1,\tilde{m}_2),
\label{sigma-cont}
\end{equation}
where $\sigma_{e^+e^-\to H}(m)$ is the total cross section for the process
$e^+e^-\to \hbox{hadrons}$ at an energy $m$ in the $e^+e^-$ rest frame, 
and $d\sigma_{\gamma_1^*\gamma_2^*}/dx$ is the cross section for 
$e^+e^-$ annihilation into two massive photons. 
$d\sigma_{\gamma_1^*\gamma_2^*}/dx$ can be related to $d\sigma/dx$ in 
Eq.~(\ref{sigma-hel-sum}):
\begin{equation}
\frac{d\sigma_{\gamma_1^*\gamma_2^*}}{dx}(\tilde{m}_1,\tilde{m}_2)=
\left(
\frac{\tilde{m}_1^2 \tilde{m}_2^2}
     {4\pi\alpha g_{V_1\gamma}g_{V_2\gamma}}
\right)^2
\frac{d\sigma}{dx}(\tilde{m}_1,\tilde{m}_2).
\label{sigma-cont2}
\end{equation}
The ranges of integration of $\tilde{m}_1$ and $\tilde{m}_2$ in
Eq.~(\ref{sigma-cont}) are determined by the physical region for the
two-meson final state: $\tilde{m}_1\leq \sqrt{s}-m_{V_i}$ for the case
of a $\rho^0$ meson and another meson $V_i$ and
$\tilde{m}_1+\tilde{m}_2\leq \sqrt{s}$ for the case of two $\rho^0$
mesons. One can recover the narrow-width approximation by writing
\begin{equation}
\sigma_{e^+e^-\to V_i}^{\rm NW}(\tilde{m})=
\left(
\frac{4\pi\alpha g_{V_i\gamma}}{m_{V_i}^2}
\right)^2
\pi \delta(\tilde{m}^2-m_{V_i}^2).
\label{narrow-width}
\end{equation}

We parametrize the contribution of the $\rho^0$ meson to 
$\sigma_{e^+e^-\to H}(\tilde{m})$, following 
Ref.~\cite{Akhmetshin:2001ig}, as
\begin{equation}
\sigma_{e^+e^-\to \rho}(\tilde{m})=\frac{8\pi\alpha^2}{3\tilde{m}^5}
p_\pi^3(\tilde{m}^2)\theta(\tilde{m}-2m_\pi)|F_\pi(\tilde{m}^2)|^2,
\end{equation}
where $m_\pi$ is the pion mass and $p_\pi(s)$ is the pion momentum:
\begin{equation}
p_\pi(s)=\sqrt{s/4-m_\pi^2}.
\end{equation}
The form factor $F_\pi(s)$, which depends on the parameters
$M_\rho$, $\Gamma_\rho$, and $\beta$, is
\begin{equation}
F_\pi(s)=
\frac{(1+\beta)^{-1}M_\rho^2(1+d\,\Gamma_\rho/M_\rho)}
{M_\rho^2-s+f(s)-iM_\rho\Gamma_\rho(s)},
\end{equation} 
where
\begin{equation}
d=\frac{3m_\pi^2}{\pi p^2_\pi(M_\rho^2)}
\log\frac{M_\rho+2p_\pi(M_\rho^2)}{2m_\pi}+\frac{M_\rho}{2\pi p_\pi(M_\rho^2)}
-\frac{m_\pi^2M_\rho}{\pi p_\pi^3(M_\rho^2)},
\end{equation}
$\Gamma_\rho(s)$ is the energy-dependent width of the $\rho$,
\begin{equation}
\Gamma_\rho(s)=\Gamma_\rho\left[\frac{p_\pi(s)}{p_\pi(M_\rho^2)}\right]^3
\left[\frac{M_\rho^2}{s}\right]^{1/2},
\end{equation}
and the real part of the correction to the denominator is given by
\begin{subequations}
\begin{eqnarray}
f(s)&=&\Gamma_\rho\frac{M_\rho^2}{p_\pi^3(M_\rho^2)}
\left\{p_\pi^2(s)[h(s)-h(M_\rho^2)]+(M_\rho^2-s)p_\pi^2(M_\rho^2)
h'(M_\rho^2)\right\},
\\
h(s)&=&\frac{2}{\pi}\frac{p_\pi(s)}{\sqrt{s}}
\log\frac{\sqrt{s}+2p_\pi(s)}{2m_\pi},
\\
h'(s)&=&\frac{dh(s)}{ds}.
\end{eqnarray}
\end{subequations}%
The parameters $M_\rho$ and $\Gamma_\rho$ are obtained from a fit to the
data for the $e^+e^-\to \pi^+\pi^-$ cross section
\cite{Akhmetshin:2003zn}: $M_\rho=775.65\pm 0.64\pm 0.50$~MeV and
$\Gamma_\rho=143.85 \pm 1.33 \pm 0.80$~MeV. (Similar results have been
obtained in Ref.~\cite{Achasov:2006vp}.) The parameter $\beta$ is not
given in Ref.~\cite{Akhmetshin:2003zn}. We infer it from the value 
$\Gamma[\rho\to e^+e^-]=7.06 \pm 0.11 \pm 0.05$~MeV that is given in 
Ref.~\cite{Akhmetshin:2003zn} and the formula
for the electronic width that is given in Ref.~\cite{Akhmetshin:2001ig}:
\begin{equation}
\Gamma[\rho\to e^+e^-]=
\frac{2\alpha^2 p_\pi^3(M_\rho^2)}{9M_\rho\Gamma_\rho}
\frac{(1+d\,\Gamma_\rho/M_\rho)^2}{(1+\beta)^2}.
\label{lept-width}
\end{equation}
Using the values for $M_\rho$ and $\Gamma_\rho$ from 
Ref.~\cite{Akhmetshin:2003zn}, we obtain $\beta=-0.0815234$.

\begin{table}[t]
\caption{\label{crosssection}%
Cross sections in units of fb for $e^+ e^- \to V_1 + V_2$ at $E_{\rm
beam}=5.29$ GeV, calculated by using the fragmentation approximation.
The uncertainties that are shown are only those that arise from the
uncertainties in the electronic widths of the vector mesons. The first
five rows are calculated in the narrow-width approximation. The last row
is calculated by taking into account the nonzero width of the $\rho$
meson, as is described in the text.}
\begin{ruledtabular}
\begin{tabular}{l|ccccc}
$V_1$ $\backslash$ $V_2$ & $\rho^0$         &   $\omega$
           &   $\phi$  &   $J/\psi$       &   $\psi(2S)$      \\
\hline
$\rho^0$
& $139.61\pm 4.82$
& $23.47\pm 0.88$
& $35.93\pm 1.29$
& $41.67\pm 1.27$
& $15.60\pm 0.46 $ \\
$\omega$
&
& $\phantom{3}0.99\pm 0.07 $
& $\phantom{3}3.02\pm 0.14 $
& $\phantom{3}3.50\pm 0.15 $
& $\phantom{3}1.31\pm 0.05  $ \\
$\phi$
&
&
& $\phantom{3}2.30\pm 0.15 $
& $\phantom{3}5.20\pm 0.21 $
& $\phantom{3}1.94\pm 0.08  $ \\
$J/\psi$
&
&
&
& $\phantom{3}2.52\pm 0.13 $
& $\phantom{3}1.81 \pm 0.06  $ \\
$\psi(2S)$
&
&
&
&
& $\phantom{3}0.32 \pm 0.02  $ \\
\hline
$\rho^0$
& $126.08\pm 4.36$
& $22.31\pm 0.84$
& $34.18\pm 1.23$
& $39.83\pm 1.22$
& $14.92\pm 0.44$ \\
\end{tabular}
\end{ruledtabular}
\end{table}

In Table~\ref{crosssection}, we give our results for the integrated cross
sections, calculated in the narrow-width approximation, for the
production of $V_1+V_2$ for the vector mesons $V_i=\rho^0$, $\omega$,
$\phi$, $J/\psi$, and $\psi(2S)$. We also give integrated cross sections
for the production of $\rho^0+V$ in which the nonzero width of the
$\rho$ has been taken into account. The uncertainties that are shown
are only those that arise from the uncertainties in the $V-\gamma$
coupling constants that are given in Table~\ref{constants}. The effect
of the nonzero width of the $\rho$ meson is to decrease the cross
sections for $\rho^0 + \rho^0$ by about 10\%, for $\rho^0 + \omega$ and
$\rho^0 + \phi$ by about 5\%, and for $\rho^0 + J/\psi$ and $\rho^0
+ \psi(2S)$ by about 4\%.

We note that the differences between the zero-width and nonzero-width
results arise mostly from the fact that the quantity \begin{equation}
I_\rho=\int_{4m_\pi^2}^\infty
d\tilde{m}^2\sigma_{e^+e^-\to\rho}(\tilde{m}) \end{equation} differs
from the coefficient of the $\delta$~function in the narrow-width
approximation of Eq.~(\ref{narrow-width}). The quantity $I_\rho$ is
about $6\%$ smaller than the coefficient of the $\delta$~function in
Eq.~(\ref{narrow-width}). (The effect of the nonzero width of the $\rho$
meson itself is to actually {\it increase} the cross sections.) The
coefficient of the $\delta$~function in Eq.~(\ref{narrow-width}) derives
from the electronic width of the $\rho$ meson, which, in turn, is
calculated in Ref.~\cite{Akhmetshin:2003zn} by using
Eq.~(\ref{lept-width}). Eq.~(\ref{lept-width}) is derived from the
vector-meson-dominance model \cite{Akhmetshin:2001ig}. Hence, the
experimental electronic width of the $\rho$ meson that is given in
Ref.~\cite{Akhmetshin:2003zn} depends on the assumptions of that model.
An alternative definition of the electronic width of the $\rho$ meson can
be obtained by equating $I_\rho$ to the coefficient of the
$\delta$~function in Eq.~(\ref{narrow-width}) and using
Eq.~(\ref{gamma-v-ee}) to relate $g_{V_i\gamma}^2$ to $\Gamma[\rho \to
e^+e^-]$. This approach leads to the result
\begin{equation}
\Gamma[\rho \to e^+e^-]=\frac{m_\rho}{12\pi^2}I_\rho.
\label{lept-width-alt}
\end{equation}
It can be shown that the definition in Eq.~(\ref{lept-width-alt}) 
differs from the one in Eq.~(\ref{lept-width}) by terms of order 
$\Gamma_\rho^2/m_\rho^2$. 

\begin{figure}
\begin{tabular}{cc}
\epsfig{file=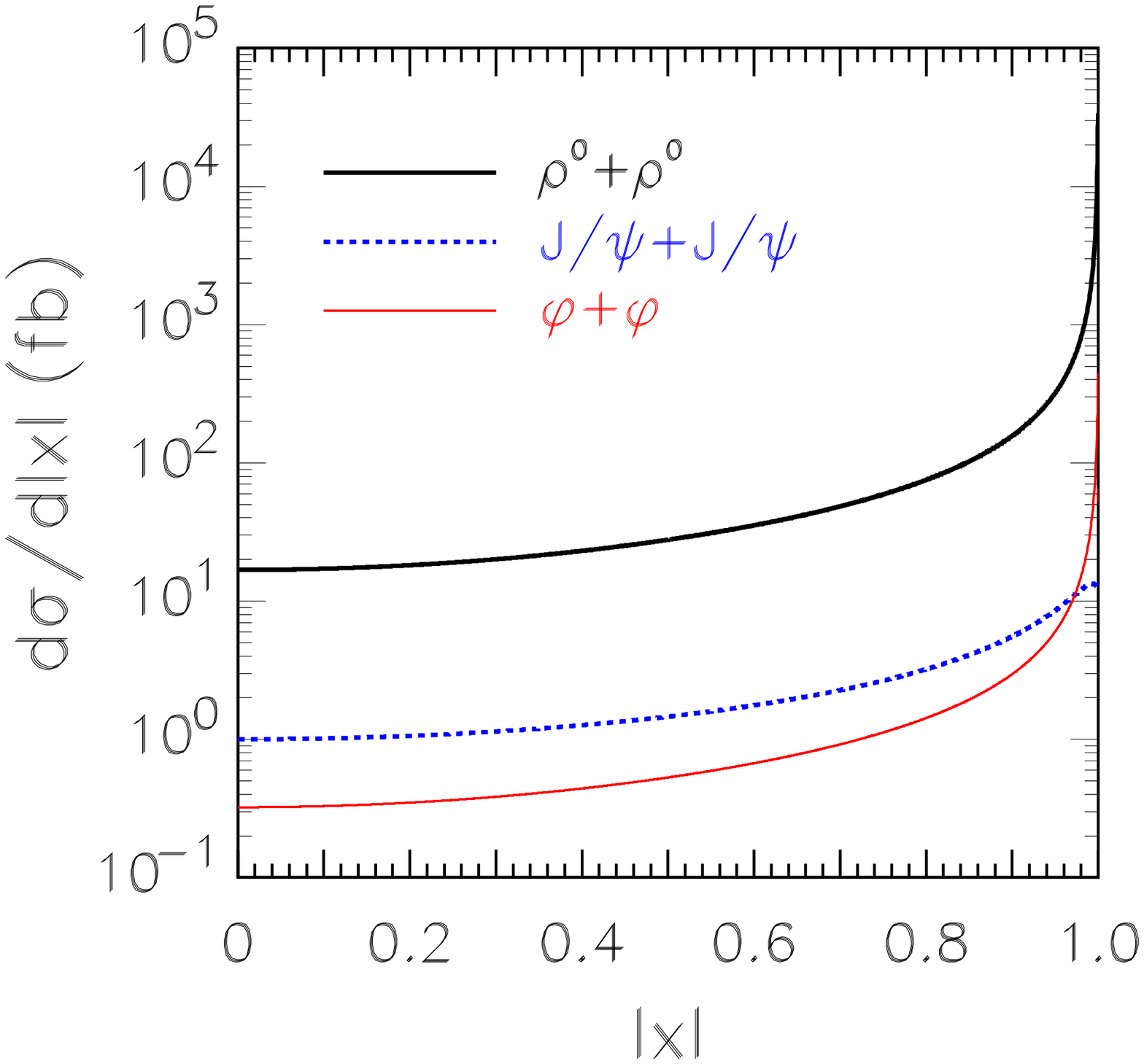,width=8.0cm}&
\epsfig{file=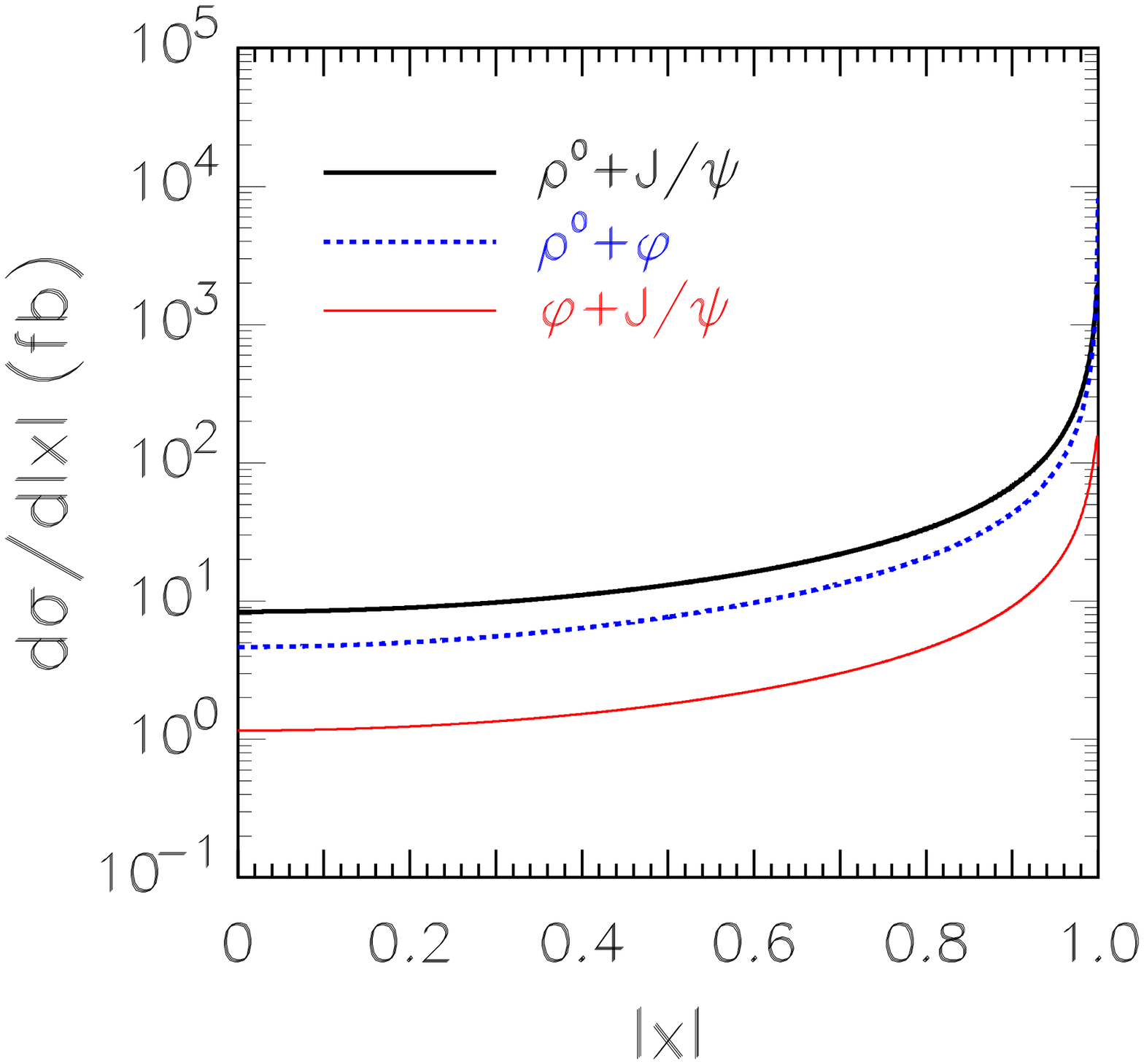,width=8.0cm}
\end{tabular}
\vspace*{2pt}
\caption{\label{fig3}%
Differential cross sections $d\sigma/d|x|$ for $e^+ e^-$ annihilation
into $V_1+V_2$ at $E_{\rm beam}= 5.29$ GeV, where $V_i= \rho^0$, $\phi$,
or $J/\psi$. The areas under the curves are the total cross sections.
}
\end{figure}
In Fig.~\ref{fig3}, we show the differential cross sections
$d\sigma/d|x|$ for the production of the identical vector mesons
$\rho^0+\rho^0$, $\phi +\phi$, and $J/\psi + J/\psi$ and for the 
production of the distinct vector mesons $\rho^0+\phi$, $\rho^0+J/\psi$,
and $\phi+J/\psi$. The differential cross sections peak sharply near
$|x|=1$ before falling to zero at the endpoint. For example, the maximum
values of the differential cross sections are $7.1 \times 10^4$ fb
at $x=0.99994$ for $\rho^0+\rho^0$ and $13.3$ fb at $x=0.993$ for
$J/\psi + J/\psi$.

We now discuss the theoretical uncertainties in the predictions for the cross
sections in Table~\ref{crosssection}. The scaling of the theoretical
uncertainties was determined in Section~\ref{ampl}. In the case of two
light vector mesons, the leading theoretical uncertainties arise from
the fragmentation approximation and from exchanges of collinear gluons
between the mesons. These uncertainties both scale as $(\Lambda_{\rm
QCD}/E_{\rm beam})^2$. If we take $\Lambda_{\rm QCD} \approx 0.5$ GeV,
then the estimated fractional uncertainty is 0.9\%. In the case
of a light vector meson and a charmonium meson, there is an additional
theoretical uncertainty that arises from the exchange of hard
gluons between the mesons that scales as $\alpha_s^2(E_{\rm beam}) \,
(m_c\Lambda_{\rm QCD}/ E_{\rm beam}^2)$. If we take $\alpha_s^2(E_{\rm
beam})=0.25$ and $m_c = 1.4$ GeV, then the estimated fractional
uncertainty is 0.2\%. In the case of two charmonium mesons, the
leading theoretical uncertainty arises from the fragmentation
approximation and scales as $m_c^2/E_{\rm beam}^2$. If we take $m_c =
1.4$ GeV, then the estimated fractional uncertainty is 7\%. This
uncertainty can be reduced by calculating the contributions to the
cross section from nonfragmentation diagrams. We discuss this
calculation in Sec.~\ref{sec:nonfrag}.

\section{Nonfragmentation corrections to the production of two
charmonia}\label{sec:nonfrag}

In this section, we calculate corrections to the production cross
sections for two charmonium mesons that arise from the nonfragmentation
contributions to the amplitudes. The contributions from the
nonfragmentation diagrams in Fig.~\ref{fig2} can be calculated using the
NRQCD factorization formalism \cite{Bodwin:1994jh}. This formalism
allows QCD radiative corrections and relativistic corrections to be
taken into account systematically. For example, the expression for the
$V-\gamma$ coupling constant defined by Eq.~(\ref{dsigmadx:VV}),
including the leading QCD radiative correction and the leading
relativistic correction, is
\begin{eqnarray}
g_{V \gamma} = 
e_c \sqrt{2 m_V}
\left(1 - \frac{8\alpha_s}{3\pi} - \frac{1}{6} \langle v^2 \rangle_V \right)
\langle V(\lambda) |\psi^\dagger \vec{\sigma} \chi | 0 \rangle 
\cdot  \vec{\epsilon}(\lambda) .
\label{g-NRQCD}
\end{eqnarray}
Here $\psi^\dagger$ and $\chi$ are the two-component Pauli operators in
the NRQCD formalism that create a heavy quark and antiquark,
respectively. $\langle v^2 \rangle_V$ is proportional to a ratio of
matrix elements of NRQCD operators between the state $V$ and the
vacuum.   The factor $e_c$ comes from the electromagnetic current in
Eq.~(\ref{current}).  The factor $\sqrt{2 m_V}$ takes into account the
difference between the standard relativistic and nonrelativistic
normalizations of the state $|V(\lambda)\rangle$.  

The NRQCD factorization formalism was used in
Refs.~\cite{Bodwin:2002fk,Bodwin:2002kk} to calculate the cross sections
for double-charmonium production from $e^+e^-$ annihilation into two
virtual photons. The calculation of the cross sections and the estimates
of theoretical errors were carried out in a way that was as close as
possible to a previous calculation of the cross sections for
double-charmonium production from $e^+e^-$ annihilation into a single
virtual photon \cite{Braaten:2002fi}. The cross sections were expressed
in terms of the coupling constants $\alpha$ and $\alpha_s$, the pole
mass $m_c$ of the charm quark, and a factor $\langle {\cal O}_1
\rangle_V$ that is   related to the NRQCD matrix element in
Eq.~(\ref{g-NRQCD}):
\begin{eqnarray}
\langle {\cal O}_1 \rangle_V = 
\frac{1}{3} \sum_\lambda 
\left| \langle V(\lambda) |\psi^\dagger \vec{\sigma} \chi | 0 \rangle \right|^2 .
\label{O-NRQCD}
\end{eqnarray}
The fragmentation terms in the cross sections in
Ref.~\cite{Bodwin:2002kk} can be recovered by replacing $g_{V\gamma}^2$
in Eq.~(\ref{dsigmadx:VV}) with $16 m_c \langle {\cal O}_1 \rangle_V/9$
and by replacing $m_V$ with $2 m_c$.  (In the calculation of
Ref.~\cite{Bodwin:2002kk}, the relative momentum of the $c$ and $\bar c$
that form the charmonium and their binding energy were neglected, and so
the invariant mass of the $c \bar c$ pair was taken to be $2 m_c$.) In
Table~II of Ref.~\cite{Bodwin:2002kk}, the cross section for
$J/\psi+J/\psi$  was given as $6.65 \pm 3.02$ fb, where the uncertainty
comes only from varying the charm quark mass over the range
$1.2~\hbox{GeV}\leq m_c\leq 1.6~\hbox{GeV}$.   This result includes both
the fragmentation diagrams in Fig.~\ref{fig1} and the nonfragmentation
diagrams in Fig.~\ref{fig2}. It does not include QCD radiative
corrections or relativistic corrections, which are known only for the
fragmentation term. As is described in Ref.~\cite{Bodwin:2002kk}, if
the known corrections to the fragmentation term in the cross section are
applied to the entire cross section, then the central value for the
$J/\psi+J/\psi$ cross section decreases to about $2$~fb.  This value is
reasonably close to the result in Table~\ref{crosssection}, but there 
are large uncertainties that arise from the uncertainty in $m_c$. 
Furthermore, this procedure suffers from the deficiency that the 
radiative and relativistic corrections to the fragmentation term in the 
cross section are not necessarily valid for the entire cross section.

A more correct procedure would be to apply the known corrections to the
fragmentation amplitude to that amplitude alone in the calculation of
Ref.~\cite{Bodwin:2002kk}. However, this approach would still be
hampered by a large uncertainty in the fragmentation amplitude that
arises from the uncertainty in $m_c$. Furthermore, there are higher-order
radiative and relativistic corrections to the fragmentation amplitude
that would not be included in this approach.

Both of these drawbacks can be eliminated by writing the fragmentation
amplitude in terms of $g_{V \gamma}$ and $m_V$ instead of 
$\langle {\cal O}_1 \rangle_V$ and $m_c$. The resulting
expression for the fragmentation term in the differential cross section
is Eq.~(\ref{dsigmadx:VV}), which has no explicit dependence on the
charm-quark mass. Note that this procedure cannot be used to account
correctly for the radiative and relativistic corrections to the
nonfragmentation amplitude. These corrections to the nonfragmentation
amplitude do not necessarily correspond to those that are contained in
$g_{V \gamma}$. Furthermore, it would be inappropriate to set $m_c$ equal
to $m_V/2$ in the nonfragmentation amplitude, since the vector-meson
mass has little to do with the propagation of a charm quark at short
distances of order $1/E_{\rm beam}$. In the case of production of
$J/\psi + \psi(2S)$, no consistent choice would be possible because
$m_{J/\psi} \ne m_{\psi(2S)}$.

We now explain precisely how we calculate improved cross sections for
two charmonium vector mesons that take into account the nonfragmentation
contributions. For each pair of vector meson helicities $\lambda_1$ and
$\lambda_2$, the cross section can be expressed as the product of a flux
factor $1/(2 s)$, the square of a T-matrix element, and a phase-space
factor $\lambda^{1/2}(1, r_1^2, r_2^2)/(8 \pi)$. We use the physical
vector meson masses in the phase-space factor. The T-matrix element is
the sum of a fragmentation amplitude and a nonfragmentation amplitude.
The cross section is the sum of a fragmentation contribution, 
a nonfragmentation contribution, and an interference contribution. 
Our expression for the fragmentation amplitude is proportional to 
$g_{V_1\gamma}g_{V_2\gamma}$, with a coefficient that is a function 
of the meson masses $m_{V_1}$ and $m_{V_2}$. Hence, 
the fragmentation contribution to the cross section is given by
Eq.~(\ref{dsigmadx:VV}). Our expression for the nonfragmentation
amplitude is proportional to $(4 m_{V_1} m_{V_2} \langle{\cal
O}\rangle_{V_1} \langle{\cal O}\rangle_{V_2})^{1/2}$, with a coefficient
that is a function of the quark mass $m_c$. The factors $(2
m_{V_i})^{1/2}$ arise from the relativistic normalizations of the meson
states. Aside from these normalization factors, the square of the
nonfragmentation contribution to the T-matrix is identical to that in
Ref.~\cite{Bodwin:2002kk}. The interference contribution to the square of the
T-matrix element depends on both the meson masses and the quark mass. We
calculate the cross sections for each helicity combination and then add
them. This approach is convenient for taking into account the difference
between the polarization vector for a meson with mass $m_{V_i}$ and the
polarization vector for a quark pair with invariant mass $2 m_c$. The
interference and nonfragmentation contributions to the cross section are
given in Eqs.~(\ref{int-hel-sum}) and (\ref{nonfrag-hel-sum}) in the 
Appendix.

Next, let us describe how we estimate the residual theoretical
uncertainties. The uncertainties are obtained by adding five theoretical
uncertainties in quadrature. The only significant uncertainties in the
fragmentation amplitudes are those that arise from the electronic widths
of the vector mesons, which enter through the overall factor of $g_{V_1
\gamma} g_{V_2 \gamma}$. These uncertainties affect the fragmentation
and interference terms in the cross sections. The most significant
uncertainties in the nonfragmentation amplitudes arise from the NRQCD
matrix elements, QCD radiative corrections, relativistic corrections,
and the charm-quark mass. These uncertainties affect the interference
and nonfragmentation contributions to the cross sections. We estimate
the error associated with the charm-quark mass by varying $m_c$ in the
interference and nonfragmentation terms in the cross section over the
range $m_c = 1.4 \pm 0.2$ GeV. Our estimates of the uncertainties in the
NRQCD matrix elements $\langle{\cal O}_1\rangle_V$ are described below.
We assume that the QCD radiative corrections to the nonfragmentation
amplitude are of relative size $\alpha_s(2m_c)=0.25$. We assume that the
relativistic corrections are of relative size $(\langle
v^2\rangle_{V_1}+\langle v^2\rangle_{V_2})/2$. Radiative corrections to
the fragmentation amplitude that involve the exchange of hard gluons
between the two mesons are suppressed as $\alpha_s^2(E_{\rm beam}) \,
(m_c^2/E_{\rm beam}^2)\approx 0.4\%$ and can be neglected. Corrections
that involve the exchange of soft gluons between the two mesons are
suppressed as $(m_cv)^4/E_{\rm beam}^4 \approx 0.04\%$ and can also be
neglected.

We compute the NRQCD matrix elements 
$\langle {\cal O}_1 \rangle_{J/\psi}$ and 
$\langle {\cal O}_1 \rangle_{\psi(2S)}$ from the electronic widths
of the charmonia by making use of the formula
\begin{equation}
\Gamma[V \to e^+e^-]=\frac{8e_c^2\pi\alpha^2}{3}
\frac{\langle{\cal 
O}_1\rangle_V}{m_{V}^2}\left(1-\frac{8}{3}\frac{\alpha_s}{\pi}
-\frac{1}{6}\langle v^2\rangle_V\right)^2,
\label{lept-width-nrqcd}
\end{equation}
which follows from Eq.~(\ref{g-NRQCD}). Note that
Eq.~(\ref{lept-width-nrqcd}) differs from the NRQCD factorization
formula that is given in Ref.~\cite{Bodwin:1994jh}: $m_c$ has been
replaced with $m_V/2$ and, consequently, the relativistic correction has
changed from $-\frac{2}{3}\langle v^2\rangle_V$ to $-\frac{1}{6}\langle
v^2\rangle_V$. The formula (\ref{lept-width-nrqcd}) is less subject to
uncertainties in the value of $\langle v^2\rangle_V$ than the standard
NRQCD formula and, through its dependence on $m_V$, resums some
corrections of higher order in $v$. In our calculation, we take $\langle
v^2\rangle_{J/\psi}=0.25\pm 0.09$ (Ref.~\cite{Bodwin:2006dn}),
$\langle v^2\rangle_{\psi(2S)}= 0.45 \pm 0.19$
(Ref.~\cite{kang}), and $\alpha_s(2m_c)\approx 0.25$. We obtain
\begin{eqnarray}
\langle{\cal O}_1\rangle_{J/\psi}&=& 0.482 \pm 0.049~\hbox{GeV}^3,
\nonumber\\
\langle{\cal O}_1\rangle_{\psi(2S)}&=& 0.335\pm 0.080~\hbox{GeV}^3.
\label{ME-values}
\end{eqnarray}
The error bars are obtained by combining in quadrature the uncertainties
from the electronic widths, the uncertainties from the values of $\langle
v^2\rangle_{V_i}$, and the estimated errors from uncalculated radiative
and relativistic corrections, which we assume to be of relative size
$\alpha_s^2$ and $\langle v^2\rangle_V^2$ in the rate, respectively.
Note that the values of the NRQCD matrix elements in
Eq.~(\ref{ME-values}) are considerably larger than those that were used
in Refs.~\cite{Bodwin:2002fk,Bodwin:2002kk}.

Our results for the cross sections for the production of $J/\psi +
J/\psi$, $J/\psi + \psi(2S)$, and $\psi(2S) + \psi(2S)$ are given in
Table~\ref{crosssectioncharmonium}. Note that the error bars for the
total cross section are less than the error bars for the fragmentation,
interference, and nonfragmentation contributions added in quadrature
because the interference uncertainties are 100\% anticorrelated with the
fragmentation and nonfragmentation uncertainties. The cross section for
$J/\psi + J/\psi$ is $1.69\pm 0.35$~fb. The central value is $33\%$
smaller than the value in the fragmentation approximation
(Table~\ref{crosssection} or the first row of
Table~\ref{crosssectioncharmonium}). The $33\%$ correction to the
fragmentation approximation is much larger than the estimate
$m_c^2/E_{\rm beam}^2\approx 7\%$ that was given at the end of
Sec.~\ref{sec:differential}. The reason for the large relative size of
this correction is that radiative and relativistic corrections to the
fragmentation contribution to the cross section, which are contained
implicitly in the quantities $g_{V_i\gamma}$ and $m_{V_i}$, decrease
the fragmentation contribution by about a factor of three from
its value without radiative and relativistic corrections.  We are able
to obtain a reasonably accurate result for the $J/\psi + J/\psi$ total
cross section, in spite of the large relative errors in the
nonfragmentation and interference terms, because the total cross section
 is dominated by the fragmentation term.
 
\begin{table}[t]
\caption{\label{crosssectioncharmonium}%
Cross sections in units of fb for $e^+ e^- \to V_1 + V_2$ at $E_{\rm
beam}=5.29$ GeV for charmonium vector mesons, calculated by including
both the fragmentation and nonfragmentation amplitudes, as is described
in the text. The four rows give the fragmentation, interference,
nonfragmentation, and total contributions to the cross sections. The
uncertainties are obtained by combining five uncertainties in
quadrature, as is described in the text.}
\begin{ruledtabular}
\begin{tabular}{l|ccc}
cross section & $J/\psi+J/\psi$ & $J/\psi+\psi(2S)$ & $\psi(2S)+\psi(2S)$
\\
\hline
fragmentation
& $ 2.52 \pm 0.13$
& $ 1.81 \pm 0.06$
& $ 0.32 \pm 0.02$
\\
interference
& \hspace{-2ex}$ -0.98 \pm 0.48$
& \hspace{-2ex}$ -1.09 \pm 0.60$
& \hspace{-2ex}$ -0.30 \pm 0.19$
\\
nonfragmentation
& $ 0.15 \pm 0.16$
& $ 0.23 \pm 0.29$
& $ 0.09 \pm 0.14$
\\
total
& $1.69 \pm 0.35$
& $0.95 \pm 0.36$
& $0.11 \pm 0.09$
\end{tabular}
\end{ruledtabular}
\end{table}

\section{Comparison with experiment and  previous theoretical
results}\label{sec:comparison}

The BaBar Collaboration has recently measured the cross sections for 
production of $\rho^0+\rho^0$ and $\rho^0+\phi$ in $e^+ e^-$ collisions at 
energy $E_{\rm beam}=5.29$ GeV~\cite{Aubert:2006we}: 
\begin{subequations}
\label{sigmaBabar}
\begin{eqnarray}
\sigma[e^+ e^- \to \rho^0+\rho^0]  &=& 
20.7 \pm 0.7_{\rm stat} \pm 2.7_{\rm syst} \ {\rm fb},
\\
\sigma[e^+ e^- \to \rho^0+\phi] &=& 
5.7 \pm 0.5_{\rm stat} \pm 0.8_{\rm syst} \ {\rm fb}.
\end{eqnarray}
\end{subequations}%
The measured cross 
sections are subject to the cuts $|\cos \theta| < 0.8$, 
$0.5~\hbox{GeV}<m_\rho<1.1~\hbox{GeV}$, and 
$1.008~\hbox{GeV}<m_\phi<1.035~\hbox{GeV}$.

In Table~\ref{crosssectionwithcut}, we give our predictions for the
cross sections that involve light mesons, integrated over the region
$|\cos \theta| < 0.8$. In Table~\ref{charmcrosssection-cut}, we give
our predictions for the charmonium-charmonium cross sections, integrated
over the same angular region. The cross sections in
Tables~\ref{crosssectionwithcut} and \ref{charmcrosssection-cut} are
much smaller than those in Table~\ref{crosssection} and
Table~\ref{crosssectioncharmonium}, respectively, because the cut on
$\theta$ excludes most of the peak near $|x|=1$. In the last row of
Table~\ref{crosssectionwithcut}, the $m_\rho$ cut is implemented. We can
calculate the effect of the $m_\phi$ cut as follows. The cross section to
produce a $\phi$ meson and one of the other vector mesons is a constant,
to within $1.5\%$, over the range
$1.008~\hbox{GeV}<m_\phi<1.035~\hbox{GeV}$. Hence, we can calculate the
fractional correction from the $m_\phi$ cut, with errors of less than
$1\%$, by calculating the integral with respect to $m_\phi^2$ of the
$\phi$ line shape over the range
$1.008~\hbox{GeV}<m_\phi<1.035~\hbox{GeV}$ and comparing it with the
integral of the line shape from threshold to infinity. Using a simple
Breit-Wigner line shape and taking the values of $m_\phi$ and
$\Gamma_\phi$ in Table~\ref{constants}, we find that the fraction 0.898
of the integral of the line shape is contained in the range
$1.008~\hbox{GeV}<m_\phi<1.035~\hbox{GeV}$. Therefore, in order to
compare the results in Table~\ref{crosssectionwithcut} with the BaBar
results, one should multiply each entry in
Table~\ref{crosssectionwithcut} by a factor 0.898 for each $\phi$ meson
in the final state. Specifically, we obtain $\sigma[e^+e^- \to
\rho^0+\phi]=5.04\pm 0.18$~fb with the mass cuts on both $m_\rho$ and
$m_\phi$.  Our predictions for $\rho^0+\rho^0$ and $\rho^0+\phi$ in
Table~\ref{crosssectionwithcut} agree with the BaBar results in
Eq.~(\ref{sigmaBabar}) to within the experimental and theoretical
uncertainties.
\begin{table}[t]
\caption{\label{crosssectionwithcut}%
Cross sections in units of fb for $e^+ e^- \to V_1 + V_2$ at $E_{\rm
beam}=5.29$ GeV, calculated using the fragmentation approximation. The
angular cut $|\cos \theta| < 0.8$, which is used in
Ref.~\cite{Aubert:2006we}, has been applied. The uncertainties shown are
only those that arise from the uncertainties in the electronic widths of
the vector mesons. The first three rows are calculated in the
narrow-width approximation. The last two rows are calculated by taking
into account the nonzero width of the $\rho$ meson, as is described in
the text. In the last row, the cut
$0.5~\hbox{GeV}<m_\rho<1.1~\hbox{GeV}$, which is used in
Ref.~\cite{Aubert:2006we}, has been applied. }
\begin{ruledtabular}
\begin{tabular}{l|ccccc}
$V_1$ $\backslash$ $V_2$ & $\rho^0$         &   $\omega$
           &   $\phi$  &   $J/\psi$       &   $\psi(2S)$      \\
\hline
$\rho^0$
& $23.30\pm 0.80$
& $3.93\pm 0.15$
& $6.43\pm 0.23$
& $10.92\pm 0.33$
& $4.44 \pm 0.13 $ \\
$\omega$
&
& $0.17\pm 0.01 $
& $0.54\pm 0.02 $
& $0.92\pm 0.04 $
& $0.37\pm 0.02 $ \\
$\phi$
&
&
& $0.44\pm 0.03 $
& $1.50\pm 0.06 $
& $0.61\pm 0.02 $ \\
\hline
$\rho^0$ (no mass cut)
& $20.50\pm 0.71$
& $3.68\pm 0.14$
& $6.03\pm 0.22$
& $10.24\pm 0.31$
& $4.17\pm 0.12$ \\
\hline
$\rho^0$ (mass cut)
& $17.71\pm 0.61$
& $3.42 \pm 0.13$
& $5.61 \pm 0.20$
& $9.52\pm 0.29$
& $3.87\pm 0.11$ \\
\end{tabular}
\end{ruledtabular}
\end{table}
\begin{table}[t]
\caption{\label{charmcrosssection-cut}%
Cross sections in units of fb for $e^+ e^- \to V_1 + V_2$ at $E_{\rm
beam}=5.29$ GeV for charmonium vector mesons, calculated by including
both the fragmentation and nonfragmentation amplitudes, as is described
in the text. The angular cut $|\cos\theta|<0.8$ has been applied. The
four rows give the fragmentation, interference, nonfragmentation, and
total  contributions to the cross sections. The uncertainties are
obtained by combining five uncertainties in quadrature, as is described
in the text.}
 \begin{ruledtabular}
 \begin{tabular}{l|ccc}
 cross section & $J/\psi+J/\psi$ & $J/\psi+\psi(2S)$ & $\psi(2S)+\psi(2S)$
 \\
 \hline
 fragmentation
 & $ 1.20 \pm 0.06$
 & $ 0.94 \pm 0.03$
 & $ 0.18 \pm 0.01$
 \\
 interference
 & \hspace{-2ex}$ -0.72 \pm 0.36$
 & \hspace{-2ex}$ -0.81 \pm 0.45$
 & \hspace{-2ex}$ -0.22 \pm 0.15$
 \\
 nonfragmentation
 & $ 0.13 \pm 0.14$
 & $ 0.20 \pm 0.25$
 & $ 0.08 \pm 0.12$
 \\
 total
 & $0.60 \pm 0.24$
 & $0.33 \pm 0.24$
 & $0.04 \pm 0.06$
 \end{tabular}
 \end{ruledtabular}
 \end{table}

Our results in Tables~\ref{crosssection} and \ref{crosssectionwithcut}
differ somewhat from those in Ref.~\cite{Davier:2006fu}. Some of the
differences arise because we are using the 2006 compilation of the
Particle Data Group \cite{PDG2006}, rather than the 2004 compilation
\cite{Eidelman:2004wy}, for the meson masses and widths. However, once
these differences in the input data are taken into account,
discrepancies still remain. The largest discrepancy is between the
results for $\rho^0 +\psi(2S)$ production with nonzero $\rho$-meson width
and $|\cos\theta|<0.8$. This discrepancy is about $27\%$.  Other
discrepancies for cross sections computed with the cut
$|\cos\theta|<0.8$ are approximately $4\%$ or less. For cross sections
computed with $|\cos\theta|<1.0$, the discrepancies are as large as
$10\%$.

The production cross sections for two vector-meson charmonium states 
were also calculated in the fragmentation and narrow-width 
approximations in Ref.~\cite{Luchinsky:2003yh}. In this paper, somewhat 
different values for the electronic widths of the charmonium states were 
used, and the resulting production cross sections are lower than ours by
about $10$--$30\%$.

The Belle Collaboration has set an upper limit on the cross 
section for $J/\psi+J/\psi$ in $e^+ e^-$ collisions at
energy $E_{\rm beam} \approx 5.29$ GeV~\cite{Abe:2004ww}: 
\begin{subequations}
\begin{eqnarray}
\sigma[e^+ e^- \to J/\psi+J/\psi] \times {\cal B}_{>2}[J/\psi] &<& 
9.1 \ {\rm fb} \hspace{1cm} (90\% \ {\rm C.L.}),
\label{sigmaBelle}
\\
\sigma[e^+ e^- \to J/\psi+\psi(2S)] \times {\cal B}_{>2}[\psi(2S)] &<& 
5.2 \ {\rm fb} \hspace{1cm} (90\% \ {\rm C.L.}),
\end{eqnarray}
\end{subequations}%
where ${\cal B}_{>2}[V]$ is the branching fraction of $V$ 
into final states with  more than two charged tracks.
These upper limits are compatible with the predictions 
in Table~\ref{crosssectioncharmonium}.
By adding up exclusive branching fractions for $J/\psi$ decays
\cite{PDG2006}, one can show that the 
branching fraction in Eq.~(\ref{sigmaBelle}) satisfies
$13\% < {\cal B}_{>2}[J/\psi] < 80\%$.
It should be possible to measure ${\cal B}_{>2}[J/\psi]$ 
and ${\cal B}_{>2}[\psi(2S)]$ at CLEOc or BESIII.
The upper limit in Eq.~(\ref{sigmaBelle}) was obtained with 
a data sample of 155 fb$^{-1}$ at or near the $\Upsilon(4S)$.  
The combined data samples of the Belle and BaBar experiments 
now exceed 1000 fb$^{-1}$.  Our prediction for the cross 
section for $e^+ e^- \to J/\psi+J/\psi$ indicates that 
there is a possibility that this process can be observed 
at the $B$ factories.

\appendix
\section{Interference and nonfragmentation contributions to the 
cross sections}

In this appendix, we give the interference and nonfragmentation
contributions to the differential cross section for $e^+ e^- \to V_1
(\lambda_1) + V_2 (\lambda_2)$, where $V_1$ and $V_2$ are charmonium
vector mesons, and $\lambda_1$ and $\lambda_2$ are their helicities. The
differential cross section for the fragmentation contribution can be
found in Eq.~(\ref{dsigmadx:VV}). The interference contribution to the
differential cross section is
\begin{eqnarray}
\frac{d \sigma^\textrm{int}}{d x} [V_1(\lambda_1) + V_2(\lambda_2)]&=&
-\frac{1024 \pi^3 \alpha^4 
     g_{V_1\gamma} g_{V_2\gamma}
    (4e_c^4m_{V_1}m_{V_2}
    \langle O \rangle_{V_1}\langle O \rangle_{V_2})^{1/2}
    \lambda^{1/2}(1,r_1^2, r_2^2)}
     {3 s^5 r^2 r_1^2 r_2^2
    (1-r_1^2-r_2^2)^2 [ 1-(1-\Delta) x^2]}\nonumber\\
&&\qquad\times F^\textrm{int}_{\lambda_1,\lambda_2}(r_1,r_2,r,x),
\label{sig-int}
\end{eqnarray}
where $r=4m_c/\sqrt{s}$. (Note that $r_1$ and $r_2$ reduce to $r/2$ in
the nonrelativistic limit.) In Eq.~(\ref{sig-int}), the numerator
factor in parentheses corresponds to the expression for $g_{V_1\gamma}
g_{V_2\gamma}$ at leading order in $\alpha_s$ and $v$
[Eq.~(\ref{g-NRQCD})]. Similar factors appear in subsequent equations
for cross sections in this Appendix. The functions
$F^\textrm{int}_{\lambda_1,\lambda_2}(r_1,r_2,r,x)$ are given by
\begin{subequations}
\label{inthelicity}
\begin{eqnarray}
F^\textrm{int}_{\pm 1, \pm 1}(r_1,r_2,r,x) &=& r^2 x^2(1-x^2) [
      (1-r_1^2 -r_2^2)-\lambda(1,r_1^2,r_2^2) ], \\
F^\textrm{int}_{\pm 1, \mp 1}(r_1,r_2,r,x) &=& (1-r_1^2 -r_2^2) (1-x^4), \\
F^\textrm{int}_ {\pm 1, 0 }(r_1,r_2,r,x) &=& 
   r r_2 [ (1-r_1^2 -r_2^2) (1-x^2)(1-2 x^2)
    +4 r_1^2 x^4 ],\\
F^\textrm{int}_{0, \pm 1}(r_1,r_2,r,x) &=& 
   r r_1 [ (1-r_1^2 -r_2^2) (1-x^2)(1-2 x^2)
    +4 r_2^2 x^4 ],\\
F^\textrm{int}_{0, 0}(r_1,r_2,r,x) &=& 4 r_1 r_2 (1+ r^2) x^2 (1-x^2). 
\end{eqnarray}
\end{subequations}%
The nonfragmentation contribution to the differential cross section is
\begin{equation}
\frac{d \sigma^\textrm{nf}}{d x} [V_1(\lambda_1) + V_2(\lambda_2)] =
\frac{8192 \pi^3 \alpha^4 
      (4e_c^4m_{V_1}m_{V_2}
      \langle O \rangle_{V_1} \langle O \rangle_{V_2}) 
      \lambda^{1/2}(1,r_1^2, r_2^2)
      F^\textrm{nf}_{\lambda_1,\lambda_2}(r,x)}
     {9 s^5 r^4 },
\end{equation}
where
\begin{subequations}
\label{nonfraghelicity}
\begin{eqnarray}
F^\textrm{nf}_{\pm 1, \pm 1}(r,x) &=& 2 r^4 x^2 (1-x^2), \\
F^\textrm{nf}_{\pm 1, \mp 1}(r,x) &=& 2 (1-x^4), \\
F^\textrm{nf}_{\pm 1, 0 }(r,x) &=& F^\textrm{nf}_{0, \pm 1}(r,x) =
    r^2 (1-3 x^2 + 4 x^4 ), \\
F^\textrm{nf}_{0, 0}(r,x) &=& 2 (1+ r^2)^2 x^2 (1-x^2). 
\end{eqnarray}
\end{subequations}%
After summing over helicity states, one finds that the interference and 
nonfragmentation contributions to the differential 
cross section become
\begin{eqnarray}
\frac{d\sigma^{\textrm{int}}}{dx}(m_{V_1},m_{V_2},m_c)
&=&
-\frac{2048\pi^3\alpha^4g_{V_1\gamma} g_{V_2\gamma}
   (4e_c^4m_{V_1}m_{V_2}
   \langle O \rangle_{V_1} \langle O \rangle_{V_2})^{1/2}
      \lambda^{1/2}(1,r_1^2,r_2^2)}
       {3s^5 r^2 r_1^2 r_2^2 (1-r_1^2-r_2^2)^2
        [1-(1-\Delta)x^2]}\nonumber\\
&&\times\bigg\{
4r r_1 r_2(r_1+r_2)
+(1-x^2)
\big[ r^2(r_1+r_2)^2(1-r_1+r_2)(1+r_1-r_2)\nonumber\\
 &&\qquad -r(r_1+r_2)(1-r_1^2+8r_1r_2-r_2^2)
+ 2(1 - r_1^2 + r_1r_2 - r_2^2)\big]\nonumber\\
&&\qquad-(1-x^2)^2[1-r(r_1+r_2)]^2
(1-r_1+r_2)(1+r_1-r_2)
\bigg\}\label{int-hel-sum}
\end{eqnarray}
and
\begin{eqnarray}
\frac{d\sigma^{\textrm{nf}}}{dx}(m_{V_1},m_{V_2},m_c)
&=&
\frac{16384\pi^3\alpha^4
      (4e_c^4m_{V_1}m_{V_2}
      \langle O\rangle_{V_1}\langle O\rangle_{V_2})
      \lambda^{1/2}(1,r_1^2,r_2^2)}
       {9s^5r^4} \nonumber\\
  &&\times \big[ 4r^2 + (1-x^2)(3r^4 -8r^2 + 5)
     - 3(1-x^2)^2(1-r^2)^2 \big].\label{nonfrag-hel-sum}
\end{eqnarray}

\begin{acknowledgments}
\noindent
We thank Michael Peskin and Kai Yi for helpful discussions.
Work by G.~T.~Bodwin in the
High Energy Physics Division at Argonne National Laboratory is supported
by the U.~S.~Department of Energy, Division of High Energy Physics, under
Contract No.~W-31-109-ENG-38. 
The work of E.~Braaten was supported in part by
the U.~S.~Department of Energy, Division of High Energy Physics, 
under grant No.~DE-FG02-91-ER4069.
The work of J.~Lee was supported by the Second Brain Korea 21 Project and
the Basic Research Program of the Korea Science and Engineering
Foundation (KOSEF) under grant No.~R01-2005-000-10089-0.
The work of C.~Yu was supported by the
Korea Research Foundation Grant funded by Korea Government 
(MOEHRD, Basic Research Promotion FUND) (KRF-2005-075-C00008).
\end{acknowledgments}


\end{document}